# Agent-Safety Evaluations as Load-Bearing Evidence: A Vendor-Neutral, Cross-Harness Reconstructability Metric

Oleg Solozobov [1]

## Abstract

Many agent-safety evaluation results are not yet load-bearing evidence: identical nominal outcomes (task success, attack success, monitor scores) may sit atop materially different evidence regimes. No vendor-neutral, runnable instrument scores reconstructability as an evaluation-validity metric: whether captured evidence can reconstruct the decision a claim depends on. This paper introduces a property-level reconstructability metric over eight decision-property classes and a cross-harness adapter emitting per-decision Evidence Sufficiency Cards backing a per-run monitor-coverage release check. It specifies a counterfactual-replay intervention protocol, implements its replayability-precondition probe, and defines a claim-evidence overclaim gap. On public and bundled traces, without new model runs, twelve-field sufficiency spans 0.458–0.833 across four inputs sharing a surface reading; replay preconditions are unmet in every scored trace. In a synthetic release-gate pair, the sufficiency gate blocks the raw variant (0.542) and passes the instrumented (0.667). Safety-evaluation claims should travel with their reconstructability vector; a reproducibility package regenerates every reported number.

## 1. Introduction

An agent-safety evaluation reports an outcome — a task succeeded, an attack was blocked, a monitor separated benign from adversarial trajectories at some area under the ROC curve (AUROC) — and that number is read as evidence for a safety claim about how the agent decides. Yet outcome and evidence are not the same object, and recent reliability work shows the two can diverge: across fourteen models on two agent benchmarks, Rabanser et al. (2026)

[1] *Corresponding author.* Affiliation: Independent Researcher (Global). E-mail address: dev404ai@gmail.com. ORCID: https://orcid.org/0009-0009-0105-7459.

report that “while rising accuracy scores on standard benchmarks suggest rapid progress, many agents still continue to fail in practice”, a discrepancy that “highlights a major limitation of current evaluations: focusing on a single metric is not enough to understand agent behavior” (Rabanser et al., 2026). A single score “ignores whether agents behave consistently across runs, withstand perturbations, fail predictably, or have bounded error severity” (Rabanser et al., 2026) — properties that live in the run, not in the headline number. This paper asks a question prior to any such metric: when an evaluation announces an outcome, did it capture enough of the run to reconstruct the decision its safety claim depends on?

Often it did not, and the deficit is structural rather than incidental. Auditing a deployed agent requires reconstructing what it did, against which policy, and on whose authority from the records the run left behind; in practice those records fall short. As Nian et al. (2026) observe, “logs are partial or absent”, and “error paths, retries, fallbacks, approvals, and inter-agent handoffs are often missing or weakly represented”, so that “even when records exist, they rarely support mechanical policy checking, and they are seldom protected against silent modification” (Nian et al., 2026). Safety evaluations inherit this shortfall: they record only what their harness was instrumented to capture. Two evaluations can therefore report an identical outcome — the same task-success rate, attack-success rate, or monitor AUROC — atop materially different evidence regimes: one retained the tool calls, permission decisions, monitor view, and policy state needed to reconstruct the decision, the other a success flag and little else. An evaluation result is called *load-bearing evidence* here when the evidence its harness retained suffices to reconstruct the decision the claim is about; much of what the field reports is not yet load-bearing in this sense, and no leaderboard number reveals which is which.

The safety-evaluation neighbourhood has named the symptoms of this deficit from several directions, but no vendor-neutral, runnable instrument yet scores whether a given evaluation claim is reconstructable from the evidence its harness recorded. Work on agent auditability distinguishes “accountability (the ability to determine compliance and assign responsibility), auditability (the system property that makes accountability possible), and auditing (the process of reconstructing behavior from trustworthy evidence)” (Nian et al., 2026), and proposes an Auditability Card that papers claiming safety should disclose — yet it stops at per-system disclosure rather than scoring whether a specific evaluation claim is reconstructable across harnesses (Nian et al., 2026). Monitorability work argues that “observability into the decision making of modern AI systems may be required to safely deploy increasingly capable agents” (Guan et al., 2025), and control work defines evaluations that “measure whether monitoring and security protocols for AI systems prevent intentionally subversive AI models from causing harm" (Bhatt et al., 2025) — each producing the nominal outcomes whose evidential sufficiency this paper interrogates, without measuring that sufficiency. The closest prior art works the same raw materials toward a different end: Zhang et al. (2025) propose "the first automated framework for annotating failed multi-agent trajectories via counterfactual replay and programmed fault injection" (Zhang et al., 2025) for failure attribution, "pinpointing the specific agent or step responsible for an error within long execution traces" (Zhang et al., 2025). Attribution presumes the failure is reconstructable and is optimised to name a culprit; the present paper's object is the prior question of evaluation validity — whether the claim is reconstructable from captured evidence at all — across successful, benign, and adversarial runs, and its instrument can return *insufficient evidence*, a verdict an attribution system is built to avoid. §2 develops this contrast.

Many agent-safety evaluation results are not yet load-bearing evidence: identical nominal outcomes — task success, attack success, monitor AUROC — can sit atop materially different evidence regimes in which the captured evidence is often insufficient to reconstruct the policy-relevant decision the safety claim is about, and reconstructability as an evaluation-validity metric — vendor-neutral, cross-harness, with a replayability-precondition probe and a claim-evidence (overclaim) gap — measures this deficiency and improves evaluation validity.

The paper makes four contributions: (1) a property-level reconstructability metric for agent-decision evaluation claims, scoring per-property recoverability in the unit interval over eight property classes — action recoverability, permission recoverability, monitor visibility, policy-state visibility, responsibility attribution, counterfactual replayability, claim sufficiency, and evaluation-awareness confounding; (2) a vendor-neutral normalization adapter that ingests agent traces from at least two evaluation harnesses into a shared Evidence Sufficiency Card, a per-decision record backing a monitor-coverage release check; (3) a counterfactual-replay intervention protocol — altered monitor visibility, permissions, policy state, or evaluation-awareness cues — with an implemented replayability-precondition probe that scores whether captured evidence satisfies the preconditions of such a replay (executed replay is the v2 protocol); and (4) a claim-evidence (overclaim) gap metric and a demonstration, on existing public and bundled traces, that inputs sharing a surface reading sit atop materially different evidence sufficiency.

This construct does not begin from scratch. The author's prior pilot introduced property-level reconstructability for agent decisions, observing that "per-property reconstructability of an agent decision already varies between regimes at this anchor scale" across six vendor SDK trace regimes (Solozobov, 2026b). That study was a single-annotator, one-anchor-per-cell descriptive cross-vendor pilot that explicitly deferred statistical testing and a production-trace corpus to later work (Solozobov, 2026b); it measured what vendor SDKs emit, not whether a safety evaluation's claim survives the evidence its harness kept. The present paper extends that descriptive measure into an evaluation-validity construct: the unit of analysis becomes the evaluation claim rather than the vendor, the regimes become evaluation harnesses rather than SDKs, and each of the eight property classes maps to a decision property that safety-evaluation claims depend upon.

Three research questions organise the study, each posed as a falsifiable hypothesis.

RQ1 (primary) is posed at the scope this version exercises — one real harness (Inspect) plus generic JSON/JSONL agent traces, generic formats being input formats rather than a second harness, with the ControlArena reader implemented but fixture-covered only — and asks whether the evidence such inputs capture exhibits per-check reconstruction gaps, at the released scorer's field level, that differ by source and format on this bounded substrate; the paper hypothesises (H1) that such gaps are present on this substrate. The eight-property, genuinely cross-harness form of the question — the eight decision properties scored across at least two real evaluation harnesses — is the v2 target. RQ1 is falsified if reconstructability is uniformly high across this substrate, with no per-check gaps by source or format.

RQ2 (primary) is likewise posed at the level this version measures and asks whether inputs sharing a nominal outcome — or, on this bounded substrate, a shared surface reading — differ in evidence sufficiency at the released scorer's field level, the §3.4 analogue of the overclaim gap; the paper hypothesises (H2) that this field-level difference is non-zero. Definition 4's G itself is the v2-protocol measurement. RQ2 is falsified if field-level evidence sufficiency does not vary across inputs sharing a nominal outcome or surface reading.

RQ3 (secondary) is posed in the precondition form this paper scores: does captured evidence satisfy the preconditions of counterfactual replay under altered monitor visibility, permissions, or policy context? Its causal form — the hypothesis (H3) that executed replay changes the reconstructability verdict for some claims, surfacing those outcome-stable yet evidence-poor — is the v2 protocol. The precondition form is falsified if preconditions are uniformly satisfied; the causal form, if replay never changes the verdict. A null result is a calibrated, publishable finding: the harnesses are already evidence-sufficient for these claims.

§2 positions the method against its neighbourhood and failure attribution. §3 specifies the metric:

property classes, unit-interval scoring, aggregation, overclaim gap, and the counterfactual-replay protocol. §4 documents the artifact: adapter, Card schema, replay probe, monitor-coverage gate, released v0 scorer. §5 describes the demonstration substrate — existing public and bundled traces, no new model runs — and the scoring procedure. §6 reports per-check gaps, the sufficiency spread across inputs sharing only a surface reading, RQ3's precondition-form answer, the monitor-coverage pair, the null observations, and the release-gate before/after. §7 states limitations and threats to validity; §8, implications for evaluation-validity reporting and adoption; §9 concludes. The metric, adapter, and scorer ship as engineering artifacts; the demonstration is exploratory, diagnostic, not pre-registered; the controlled validation matrix with executed replay is the v2 protocol.

# 2. Related Work and Gap

The method advanced here sits inside a fast-moving neighbourhood of agent-safety evaluation, control, monitorability, and auditability work, inheriting much of its raw material — long-horizon agent traces, counterfactual replay, post-hoc operational evidence. What distinguishes the present paper is not the material but the question asked of it: prior work measures agent behaviour, scores a monitor's catch rate, attributes the failure-inducing agent or step, or gates the agent's own completion claim, whereas the question here is whether the evidence an evaluation captured suffices to reconstruct the decision its safety claim depends on. This narrative survey — non-exhaustive, no systematic database search — positions the method against its two closest neighbours, then four thematic clusters, and closes with the gap the field has named but not yet closed. The neighbourhood is preprint-heavy: most cited works are arXiv preprints or framework documentation, reflecting the field's recency, with peer-reviewed versions expected to supersede several entries.

The closest prior art, AgenTracer, works the same raw materials — counterfactual replay over long-horizon multi-agent traces — toward a different object, and the contrast fixes this paper's wedge. AgenTracer is "the first automated framework for annotating failed multi-agent trajectories via counterfactual replay and programmed fault injection" (Zhang et al., 2025), built for "pinpointing the specific agent or step responsible for an error within long execution traces" (Zhang et al., 2025). The overlap with the present method is real: both are organised around re-running recorded decisions under altered conditions — AgenTracer executes such replays, while this paper specifies the intervention protocol and scores its preconditions — and both treat the durable trace as post-hoc operational evidence, reasoning over execution histories too long for manual inspection. The divergence is in the object of study. Failure attribution presumes the failure is reconstructable from the trace and is optimised to return a culprit. The object here is the prior question of evaluation validity — whether the claim is reconstructable from the captured evidence at all — and it ranges over successful, benign, and adversarial runs rather than only failed ones. Most consequentially, the present instrument can return *insufficient evidence*: a verdict that an attribution system, built to produce an answer, is structured to avoid. Where attribution asks "which step failed?", this paper asks "could the evidence the harness kept even answer that?"

A second head-to-head, EviBound, shares the gate posture — no claim credited without machine-checkable evidence — but audits a different actor. Its governing rule, "Treat every claim as unverified until it is proven with machine-checkable artifacts" (Chen, 2025), swaps trust in agent assertions for queries against external artifact stores, and the evidence it verifies is the agent's own execution record: "queryable execution identifiers (e.g., MLflow run id with FINISHED status), artifact files in standard locations, metrics files, and execution logs" (Chen, 2025). The distrusted party is the agent; the gated object, its self-reported task completion. This paper's instrument is therefore not EviBound for safety: it audits the monitor rather than the agent — whether an identified safety monitor's observation of the execution is preserved and sufficient (§4.5) — and ranges over an evaluation's claims about a run, not an agent's reports about its own task.

The first cluster, sabotage and control evaluation, manufactures the adversarial outcomes whose evidential sufficiency this paper interrogates. Control evaluations "measure whether monitoring and security protocols for AI systems prevent intentionally subversive AI models from causing harm" (Bhatt et al., 2025). Sabotage-capability suites test whether frontier models could covertly subvert oversight (Benton et al., 2024), and SHADE-Arena pits harmful side-task completion against LLM-monitor suspicion (Kutasov et al., 2025). Ctrl-Z blocks or incriminates subversive agents by resampling suspicious actions (Bhatt et al., 2025), and CTRL-ALT-DECEIT extends sabotage evaluation to agentic ML engineering (Ward et al., 2025). Each reports a nominal outcome — an attack-success rate, a safety-versus-usefulness frontier — and this paper asks whether the recorded run retains the permission decisions, monitor view, and policy state needed to reconstruct the decision it summarises.

The second cluster, monitorability, asks whether a monitor can see enough of a run to judge it, and so motivates the monitor-visibility property class directly. The field treats "observability into the decision making of modern AI systems" as something that "may be required to safely deploy increasingly capable agents" (Guan et al., 2025), and reports that "CoT monitoring is more effective than action-only monitoring in practical settings" (Guan et al., 2025) — direct evidence that an identical nominal monitor score means different things depending on what the monitor was shown, the very confound this paper's metric records. Studies of how information access shapes a monitor's ability to detect sabotage (Arike et al., 2026) and operational monitorability and eval-integrity prototypes (METR, 2026) round out the cluster. MonitoringBench red-teams the detection side, reading each monitor's catch rate on adversarially refined attack trajectories against a calibrated suspicion threshold, and accounts for score coverage, computing over valid-score sets (Jotautaitė et al., 2026); the monitor-coverage check of §4.5 recasts that aggregate accounting as a per-run, claim-conditioned release test, so it complements rather than competes with the instrument here. None scores, per evaluation claim, whether the monitor's view was preserved well enough to reconstruct it.

The third cluster studies deception, scheming, sandbagging, and evaluation-awareness — whether agents misbehave, and whether knowing they are tested changes that. Frontier models are capable of in-context scheming, and the evidence that distinguishes deliberate from accidental misbehaviour lives in the reasoning trace: "analysis of the models' chains-of-thought reveals that models explicitly reason about these deceptive strategies, providing evidence that the scheming behavior is not accidental" (Meinke et al., 2024). Evaluations of stealth and situational awareness frame their results as evidence in a structured argument — they "present a safety case (Clymer et al., 2024) for scheming inability…, making a high-level argument that models need basic situational awareness and top human-level stealth to cause severe harm via scheming" (Phuong et al., 2025). Language models can strategically underperform on evaluations, prompted or password-locked to target specific scores (Weij et al., 2024), and in-context environments can induce evaluation-awareness (Chaudhary, 2026). Models can sandbag covertly against chain-of-thought monitoring (Li & Siegel, 2025), and blind auditing games test whether investigators uncover a model's hidden objective (Marks et al., 2025). Each safety claim is only as good as the cues the trace retained: if conduct under test was shaped by awareness of the test and those cues went unrecorded, the nominal outcome describes evaluation conditions, not deployment — which is why this paper makes evaluation-awareness confounding a scored property.

The fourth cluster turns directly to auditability, traceability, logging, and safety cases — the substrate on which the others rest, and the one the present construct most directly extends. Auditable Agents distinguishes "accountability (the ability to determine compliance and assign responsibility), auditability (the system property that makes accountability possible), and auditing (the process of reconstructing behavior from trustworthy evidence)" (Nian et al., 2026), the vocabulary this paper inherits. Trace-issue localisation contributes human-annotated agentic traces and shows the difficulty of automated reconstruction: "our evaluations reveal that modern long context LLMs perform poorly at trace debugging, with the best GEMINI-2.5-PRO model scoring a mere 11% on TRAIL" (Deshpande et al., 2025), a result that motivates a deterministic, schema-based sufficiency check rather than another model-in-the-loop judge. Structured log-analysis guidance (Dubois et al., 2026), evaluations-based safety cases for scheming (Balesni et al., 2024), and agentic red-teaming environments (Tien et al., 2026) add methodology. Evaluation of whether models would sabotage AI-safety research (Kirk et al., 2026) completes the cluster. Operationally, this neighbourhood runs on open evaluation harnesses — the UK AI Security Institute's Inspect (UK AI Security Institute (AISI), 2024) and ControlArena (UK AI Security Institute (AISI); Redwood Research, 2025), and behaviour-trace corpora such as MALT (METR, 2025) — Inspect and ControlArena being the integration targets of the cross-harness adapter in §4, exercised on Inspect in §5. LinuxArena carries the control setting into live production software, defining action-level monitor coverage (Tracy et al., 2026).

Read together, these lines name the symptom the present metric measures, yet each stops short of a vendor-neutral, runnable instrument that scores whether a specific evaluation claim is reconstructable from the evidence its harness recorded. Auditable Agents states that in deployed systems "logs are partial or absent… and… even when records exist, they rarely support mechanical policy checking" (Nian et al., 2026), and proposes per-system disclosure via an Auditability Card rather than a cross-harness score tied to a particular claim (Nian et al., 2026). Its metrics — Action Coverage Rate and Record Fidelity, scoring "whether policy-relevant actions appear in the record at all" and "whether enough fields survive for each recorded action to support reconstruction" (Nian et al., 2026), with Lifecycle Phase Coverage and Gap Burden extending the accounting to execution phases (Nian et al., 2026) — measure record completeness, descriptive and post hoc, a different quantity from the monitor-observation sufficiency gated in §4.5. At governance altitude, ADAS compiles "statutory and regulatory obligations into machine-executable deployment gates" (Saparning, 2026), runs a literal "Evidence Sufficiency Check, verifying that all mandated evidence types are present, correctly hashed, and internally consistent" (Saparning, 2026), and issues a cryptographically verifiable per-deployment certificate (Saparning, 2026); despite the shared phrase, it authorises a model deployment under law — the instrument here checks one agent run, at trace level, for whether the monitor's observation of the execution was preserved. The control line flags the missing tooling as an open problem: "future work needs to investigate AI-assisted auditing techniques that require less involvement and lower reliability from humans" (Bhatt et al., 2025). The monitorability line establishes that the catch rate depends on what the monitor saw (Guan et al., 2025) without scoring that dependence per claim. The scheming line rests its conclusions on structured safety cases (Phuong et al., 2025) whose evidential sufficiency is asserted rather than measured. And the log-analysis line poses the open question outright — "what are the best logging and log analysis methodologies for identifying abstract behaviours such as 'evaluation awareness'?" (Dubois et al., 2026). The author's prior reconstructability pilot answered a narrower version of this for vendor SDKs, finding that "per-property reconstructability of an agent decision already varies between regimes at this anchor scale" (Solozobov, 2026b), but as a descriptive cross-vendor study it measured what SDKs emit, not whether a safety evaluation's claim survives the evidence its harness kept (Solozobov, 2026b). The territory is active enough to have drawn a dedicated survey of evidence tracing and execution provenance in LLM agents (Wang et al., 2026), yet no prior work provides a vendor-neutral, cross-harness, per-run, claim-conditioned instrument that admits *insufficient evidence* as a verdict and scores monitor-observation sufficiency — did an identified monitor score the span on which the claim rests — and that runnable metric is this paper's wedge.

# 3. Method

This section specifies the instrument the paper contributes: a vendor-neutral, model-agnostic metric that scores, per decision, whether the evidence an evaluation harness recorded suffices to reconstruct the decision its safety claim depends on. The metric is definitional here, so no scores or outcomes are reported; the normalization adapter and the per-decision Evidence Sufficiency Card schema that operationalise it are deferred to §4. Throughout, the unit of analysis is the *evaluation claim* — an assertion a safety evaluation makes about an agent decision on the strength of a recorded run — and a claim is *reconstructable* to the degree that the decision properties it depends on can be recovered from the captured evidence. The metric measures that degree, neither adjudicating whether behaviour was safe nor ranking harnesses or vendors. The construction proceeds through the eight property classes (§3.1), per-property scoring (§3.2), aggregation (§3.3), the claim-evidence gap (§3.4), and the replay intervention protocol with its replayability-precondition probe (§3.5).

## 3.1. The eight property classes

A safety-evaluation claim about an agent decision depends on a small, fixed set of decision properties, each reconstructed from a distinct kind of recorded evidence; the metric measures reconstructability over eight such property classes — action recoverability, permission recoverability, monitor visibility, policy-state visibility, responsibility attribution, counterfactual replayability, claim sufficiency, and evaluation-awareness confounding — treating the absence of any one as a bounded gap, not a global failure. The eight classes extend the prior reconstructability pilot, in which each Decision Event Schema property of a recorded decision was classified as one of "fully fillable (anchor fragment present and complete), partially fillable (fragments exist but evidence is split or incomplete), structurally unfillable (no fragment of the required type appears), opaque (the property class is observable in principle but the evidence is not externally inspectable…)" across six vendor SDK trace regimes (Solozobov, 2026b). They cross-walk the auditability literature, which "define[s] five dimensions of agent auditability, i.e., action recoverability, lifecycle coverage, policy checkability, responsibility attribution, and evidence integrity" as jointly necessary for a defensible audit verdict (Nian et al., 2026): the present paper refactors that frame to the evaluation-claim unit: action recoverability and responsibility attribution carry over; policy checkability splits into its evidence requirements (permission recoverability, policy-state visibility); lifecycle coverage generalises to what the monitor saw and whether the decision point can be re-entered (monitor visibility, counterfactual replayability); and two classes load-bearing for evaluation validity but not foregrounded by a per-system audit are added (claim sufficiency, evaluation-awareness confounding).

**Definition 1 (property classes).** For an evaluation claim about a recorded agent decision, each property class pairs a decision property with the evidence that reconstructs it and the claim it underwrites: (i) *action recoverability* — the tool calls, requests, and state mutations issued, from the action record; underwrites any task- or attack-success claim. (ii) *Permission recoverability* — the authorization decision in force when the action executed, from the permission record; underwrites any allow/block/escalate claim. (iii) *Monitor visibility* — what the monitor was shown (actions, chain-of-thought (CoT), both, or a filtered view), from the monitor-input record; underwrites any monitor-score claim. (iv) *Policy-state visibility* — the policy version and rollout or context state the decision was taken against, from the policy-state record; underwrites any compliance claim. (v) *Responsibility attribution* — the agent, step, sub-agent, or human-approval node the decision is attributable to, from the attribution chain; underwrites any claim about who or what caused an outcome. (vi) *Counterfactual replayability*

— whether the decision could be re-run under altered conditions, from the presence of replay preconditions and handles; underwrites any claim that an outcome was robust rather than incidental. (vii) *Claim sufficiency* — whether the recorded evidence is jointly adequate to decide the specific claim, not merely to describe the run; underwrites any safety-case claim. (viii) *Evaluation-awareness confounding* — whether the trace records the cues (test framing, sandbagging signals, situational-awareness prompts) that would let an analyst judge whether behaviour was confounded by the agent knowing it was under evaluation; underwrites any behavioural safety claim. Claim sufficiency carries the evaluation-validity reading most explicitly, because the auditability literature needs a third verdict: a structural policy checked against an audit record "has three possible outcomes: comply, violate, or ⊥ (undecidable from the record)", and it can be "impossible to decide because the record omits a required field" (Nian et al., 2026); claim sufficiency is the property that lets the metric return *insufficient evidence* rather than a forced verdict.

These eight classes are the design-level construct; the released v0 scorer operationalises them as twelve per-card evidence checks, F01-F12 (mapping in §4.4).

## 3.2. Per-property reconstructability scoring in $[0, 1]$

The metric scores each property's reconstructability on the closed unit interval as the proportion of that property's required evidence elements that the captured trace both contains and binds to the decision under audit, recovering the pilot's discrete labels as the endpoints and midpoint of a unit-interval-valued scale rather than replacing them. The prior pilot scored reconstructability discretely, "mapping fully fillable to 1.0, partially fillable to 0.5, structurally unfillable to 0.0, opaque to 0.0" for each property (Solozobov, 2026b). That three-valued mapping suffices when a property is reconstructed from a single anchor fragment; the evaluation-harness setting routinely splits a property across several recorded elements — a permission decision may record the verdict but not the policy it was checked against — so the discrete mapping is generalised here to a unit-interval-valued one — a finite fractional grid for a fixed required-evidence set — of which the pilot's scheme is the special case. The generalisation is the paper's contribution, not derived from any external source.

**Definition 2 (per-property reconstructability score).** Fix an evaluation claim and a decision under audit. For property class p, let its *required evidence set* $R_p$ be the finite set of evidence elements a fully reconstructed instance of p comprises, with membership fixed per class in §4. Let present(p) be the subset of $R_p$ the captured trace contains and binds to this decision through the Evidence Sufficiency Card (§4). The per-property reconstructability score is the recovered proportion $r_p = |\text{present(p)}| / |R_p|$, equal to 1 when every required element is present and bound and 0 when none is. When $R_p$ is a single element, $r_p$ reduces to the pilot's 1.0/0.0 endpoints, and the half-recovered two-element case reproduces its 0.5 midpoint. An element observable in principle but not externally inspectable — the *opaque* case, paradigmatically a model's internal reasoning — counts as absent for scoring but is flagged distinctly from a *structurally absent* element on the Card. Membership is declared, not inferred, so $r_p$ is deterministic in trace and schema, computed by counting present against required fields with no learned parameters and no model in the loop.

## 3.3. Aggregation: the reconstructability vector and scalar

The metric aggregates the eight per-property scores into a per-decision reconstructability vector, treats that vector as the primary reportable quantity, and derives any scalar summary as an explicit, weakest-link-sensitive function so that no single number can mask a wholly unreconstructable property.

**Definition 3 (reconstructability vector and aggregate).** For a decision under audit, the *reconstructability vector* is the ordered tuple r = $(r_1, \ldots, r_8)$ of the eight per-property scores of Definition 2, in the fixed order of Definition 1; it is the metric's primary output, reported in full; two scalar summaries derive from it as adjuncts. The *bottleneck score* is the minimum over the claim-required properties, $R_{min} = min$ over p in R of $r_p$, where R is the subset of classes the specific claim depends on; the minimum is the headline scalar because a claim is no more reconstructable than its least-reconstructable required property. The *coverage score* is an optionally weighted mean, $R_{mean} = sum$ over p of $w_p \; r_p$ with weights summing to one and uniform $w_p = 1/8$ by default; it summarises overall coverage but can be high while a required property is zero, and is therefore never reported without the vector. The minimum is chosen over a product or sum because it is non-compensatory in the same way the evidence is: the auditability literature observes that "a single field omitted from the record schema can render an entire class of policies uncheckable, regardless of how many events are logged" (Nian et al., 2026), so one zero-scored required property drives the headline scalar to zero however richly the others are recovered. Non-uniform weights are a declared per-claim parameter, not a tuned quantity.

### 3.4. The claim-evidence (overclaim) gap

The claim-evidence (overclaim) gap is defined, for a pair of evidence regimes that report the same nominal outcome, as the divergence between their reconstructability under that shared outcome, expressed as a plain difference rather than an information-theoretic quantity, so that a non-zero gap says exactly how much less evidence the same headline number rests on in the weaker regime than in the stronger. The paper's central claim (§1) is that "identical nominal outcomes — task success, attack success, monitor AUROC — can sit atop materially different evidence regimes"; the overclaim gap makes that divergence measurable for a matched pair, anchored in the prior finding that per-property reconstructability already varies across regimes (Solozobov, 2026b), now lifted from vendor SDK regimes to evaluation harnesses and conditioned on a shared outcome.

**Definition 4 (claim-evidence (overclaim) gap).** Let A and B be two evidence regimes — two harnesses, two runtime configurations, or a rich-evidence versus an action-only recording of the same run — that report the same nominal outcome o for a claim (the same task- or attack-success flag, or monitor scores within a stated tolerance). Let $R(A)$ and $R(B)$ be their bottleneck reconstructability scores for that claim under Definition 3. The *scalar overclaim gap* is the absolute difference G(A, B) = $|R(A) - R(B)|$, a number in $[0, 1]$, approaching one as one regime supports the outcome with near-complete evidence and the other with almost none. The *property-resolved overclaim gap* is the per-property vector of absolute differences $g_p = |r_p(A) - r_p(B)|$, reported alongside G to attribute the gap to the specific properties on which the regimes diverge. The gap is defined only between regimes sharing the nominal outcome o, and its semantics are explicit: G is a symmetric comparative divergence between the two regimes' evidence sufficiency, invariant under exchanging A and B; its overclaim reading is comparative, not absolute: the weaker regime — the smaller of $R(A)$ and $R(B)$ — supports the same nominal outcome with evidence poorer by exactly G than the stronger regime's, while how far either regime's evidence falls short of the claim's own warrant is a separate quantity the definition does not fix; and a gap of zero asserts equal evidence sufficiency, not adequacy of either regime.

So defined, the gap is diagnostic, not normative: it reports how unequally two regimes' evidence supports a shared nominal outcome, and returns zero — a calibrated null — for equally evidence-sufficient regimes. Released v0 cards support a field-level analogue of G rather than G itself: each card carries a sufficiency_score over its twelve field statuses and a blocking_gaps list naming the fields whose partial or missing status holds the verdict below sufficient-for-named-decision, and for a matched-outcome pair the analogue is the difference of the two sufficiency scores, resolved to the blocking-gap entries on which the cards differ. The analogue approximates but does not instantiate Definition 4: sufficiency_score is a mean-like scalar over twelve field checks, not the bottleneck $R_{min}$ over the claim-required subset, and no equivalence between the two aggregations is established; measuring Definition 4's G as defined is the v2 protocol.

### 3.5. The replay intervention protocol and the replayability-precondition probe

The replay component is specified at two deliberately separated levels: an intervention protocol fixing which conditions of a recorded decision an executed counterfactual replay would alter — the monitor-visibility regime (actions-only, CoT-only, CoT-plus-actions, or filtered), the permission outcome (granted versus denied), the policy or rollout state, and inserted or removed evaluation-awareness cues — and a deterministic replayability-precondition probe, the level this paper implements and scores. The probe re-runs nothing: it checks, per recorded decision, whether the captured evidence satisfies six preconditions of a well-defined counterfactual replay: (i) model binding resolved — exact model identity with a version or checkpoint reference; (ii) decision context reconstructable — full input state, tool schemas, and relevant memory or policy state; (iii) determinism controls — seed or sampling parameters, or acknowledged nondeterminism bounds; (iv) the monitor-view specification present and (v) the permission and policy state present, since a condition never recorded cannot be counterfactually altered; and (vi) an environment and tool side-effect re-creation specification sufficient to re-enter the decision point. Per decision the probe emits a replay verdict — possible, degraded, or impossible — plus a missing-evidence list naming the unmet preconditions; the counterfactual-replayability score remains a Definition 2 proportion in $[0, 1]$.

This split restates RQ3 in the form the captured evidence can answer: its precondition form — for how many recorded decisions are replay preconditions unmet, leaving outcome-stable claims evidence-poor already at the precondition level — is posed here, while its causal form, whether executing the protocol changes a reconstructability verdict, is explicitly labelled the v2 protocol, not silently dropped. The anticipated objection — the protocol is specified but never executed — inverts the instrument's logic: where evidence fails the preconditions, executed replay is not merely out of scope but undefined, and the probe converts that impossibility into a measured, per-decision result. Whether replay is possible from captured evidence is a presence question;

whether a claim survives replay is a quality question and travels with the v2 protocol. §4 specifies the Card that populates present(p), the required sets, matched-outcome regimes, and precondition verdicts.

# 4. Artifact: Cross-Harness Adapter and the Evidence Sufficiency Card

This section documents the artifact operationalising the §3 metric: a vendor-neutral normalization adapter, the per-decision Evidence Sufficiency Card it emits, the replayability-precondition probe that populates the Card's replay fields, the released v0 scorer implementing the Card as twelve evidence checks, and a trace-level monitor-coverage gate on the same presence machinery. The Card is where §3's machinery becomes mechanical: it determines present(p) per property class and yields $r_p$ and the bottleneck and overclaim quantities of Definitions 3 and 4. The adapter, Card schema, and eight required-evidence sets are the paper's design contribution — specified, not measured; the concrete verdicts quoted in §4.4-§4.5 are fixed demonstration fixtures shipped with the released tool, not study results.

## 4.1. The cross-harness normalization adapter

The adapter is a vendor-neutral component that ingests agent traces from at least two evaluation harnesses and emits one Evidence Sufficiency Card per decision under audit, normalising by mapping each harness's native trace records onto a shared set of Card fields rather than adopting any one harness's format. Its design follows established precedent for heterogeneous telemetry in a single schema: the OpenTelemetry GenAI conventions define agent spans that "extend and override the semantic conventions for Gen AI Spans" (OpenTelemetry Project (CNCF), 2025), and reconcile divergent vendor formats through a discriminator field: the gen_ai.provider.name attribute "acts as a discriminator that identifies the GenAI telemetry format flavor specific to that provider" (OpenTelemetry Project (CNCF), 2025). A harness trace is, concretely, a stream of typed records — in the OpenAI Agents SDK each run carries "span_data, which is information about the Span" (OpenAI, 2025) — and the adapter extends the property-level normalization the author's prior pilot applied across six vendor SDK regimes, where a per-regime adapter parsed native records into one property schema scored deterministically from the populated fields (Solozobov, 2026b). AgentTrace instruments agents at runtime to emit structured logs (AlSayyad et al., 2026); the adapter instruments nothing, normalising post-hoc what harnesses already emitted.

Vendor-neutrality follows from the direction of the mapping: each harness keeps its native trace, and a per-harness ingest module routes that harness's record types to the Card fields they evidence — an action record to the action fields, the monitor's input transcript to the monitor-view field. Adding a harness adds an ingest module, not a change to the Card schema or the metric: the shared schema is the fixed contract, the per-harness mappings the only vendor-specific surface. The principle claims nothing about any harness's internals: the two integration targets are the UK AI Security Institute's Inspect (UK AI Security Institute (AISI), 2024) and ControlArena (UK AI Security Institute (AISI); Redwood Research, 2025), with the Inspect field mappings exercised in §5 rather than specified here (the ControlArena mappings are implemented and covered by released tests; §5.1).

Binding is what makes present(p) deterministic: each evidence element the adapter writes is bound to the decision under audit by a triple of trace identifiers — the decision identifier, the agent step, and the originating tool call — so an element counts toward present(p) only when present in the trace and attached to this decision, as Definition 2 requires. An element found elsewhere in the run but unbindable to the decision is recorded as missing, not silently credited; an element observable in principle but not externally inspectable is bound yet flagged opaque rather than absent. Because the required-evidence sets are fixed in §4.2 and the binding rule is mechanical, $r_p$ is a pure function of the populated Card with no model in the loop.

## 4.2. The Evidence Sufficiency Card

**Definition 5 (Evidence Sufficiency Card).** The Evidence Sufficiency Card is the per-decision JSON record the adapter emits; it refactors the record the author's prior pilot scored — a per-decision evidence record whose fields a deterministic, schema-based stage classified without a model in the loop (Solozobov, 2026b) — from the vendor-SDK setting to the evaluation-claim unit. The schema identifies the decision; records the monitor regime and output, the authorization, the task-level claims under audit, and replay-precondition state; and carries the metric's outputs with a per-property *reconstructability* sub-object holding the eight scores $r_p$ of Definition 2 and, per score, whether an absent element is structural or opaque. Five fields warrant note: *monitor_view* records the regime the monitor was shown — actions-only, CoT-only, CoT-plus-actions, or filtered — so a *monitor_score* is never recorded without its view; *missing_evidence* enumerates the required elements not present-and-bound per property; *claim_evidence_gap_score* carries the scalar overclaim gap of Definition 4 when the Card is one half of a matched-outcome pair; *replay_preconditions* carries the §3.5 probe verdict — possible, degraded, or impossible — with the unmet preconditions listed; and *counterfactuals_run* enumerates executed replays and stays an honest zero (empty) in this release: the Card never asserts "replayable", only that preconditions are satisfied or unmet. The schema is the fixed contract the adapter targets; the field set does not vary by harness.

```
{
  "decision_id": "string",
  "agent_step_id": "string",
  "tool_call_id": "string",
  "policy_version": "string",
  "monitor_view": "actions_only | cot_only | cot_and_actions | filtered",
  "monitor_score": "number | null",
  "permission_decision": "allow | block | escalate | none",
  "side_task_claim": "string | null",
  "main_task_claim": "string | null",
  "replay_preconditions": {
    "verdict": "possible | degraded | impossible",
    "unmet": ["precondition"]
  },
  "counterfactuals_run": ["string"],
  "missing_evidence": {"property_class": ["required_element"]},
  "claim_evidence_gap_score": "number | null",
  "reconstructability": {
    "action_recoverability": "PropertyScore",
    "permission_recoverability": "PropertyScore",
    "monitor_visibility": "PropertyScore",
    "policy_state_visibility": "PropertyScore",
    "responsibility_attribution": "PropertyScore",
    "counterfactual_replayability": "PropertyScore",
    "claim_sufficiency": "PropertyScore",
    "evaluation_awareness_confounding": "PropertyScore"
  },
  "$defs": {
    "PropertyScore": {"score": "number", "absence": "none | structural | opaque"}
  }
}
```

Definition 2 deferred each $R_p$'s membership to this section; Definition 6 supplies it, declared rather than inferred, so the denominator of every $r_p$ is a fixed, inspectable quantity.

**Definition 6 (the eight required-evidence sets).** For each property class p, the required-evidence set $R_p$ is fixed as follows, and present(p) is the subset of $R_p$ present-and-bound on the Card. (i) *Action recoverability*: the tool or function called, its arguments, the returned result or error, and the resulting state mutation. (ii) *Permission recoverability*: the requested action, the authorization verdict, the policy or rule invoked, and the principal it was granted on the authority of. (iii) *Monitor visibility*: the monitor-input view label, the exact content shown to the monitor, and the monitor's emitted score or verdict. (iv) *Policy-state visibility*: the policy version, the rollout or context state in force, and the binding of that state to the decision timestamp. (v) *Responsibility attribution*: the executing agent or sub-agent, the originating step, and any upstream delegation or human-approval node in the chain. (vi) *Counterfactual replayability*: a record of the decision inputs sufficient to re-enter the decision point, the seed or sampling state or acknowledged nondeterminism bounds, and a replay handle identifying the run — the evidence the §3.5 preconditions test. (vii) *Claim sufficiency*: the explicit claim under audit, the property classes it depends on, and the present-and-bound subset of each of those classes' required evidence. (viii) *Evaluation-awareness confounding*: the test-framing cues in context, any situational-awareness or sandbagging signals in the trace, and the reasoning record that would let an analyst judge whether the cues shaped conduct. The opaque case (vii–viii most often, but any class in principle) marks an element observable in principle but not externally inspectable as absent for scoring and opaque for remediation. These sets are the author's design and the central calibration target; their correctness is asserted, not empirically validated, and §7 treats their revisability as a threat to validity. Membership is unconditional per class in this release: the schema does not yet encode conditional applicability — a state-mutation element for a read-only action, a delegation node for a run with no delegation — so an inapplicable element scores as absent rather than not-applicable, conservatively depressing $r_p$; an applicability mask fixing an effective per-decision denominator is a calibration-stage refinement.

### 4.3. The counterfactual-replay probe: protocol specification, precondition scoring

In this release the counterfactual-replay probe is a precondition scorer, not an executor: it evaluates the six §3.5 preconditions against each recorded decision's captured evidence, writing the result to the Card's *replay_preconditions* sub-verdict while *counterfactuals_run* stays an honest zero: no replay is executed. The protocol whose preconditions it scores is fixed in §3.5 — altered monitor visibility, permissions, policy or rollout state, and evaluation-awareness cues. Executing that protocol, each replayed condition yielding a fresh Card, is the v2 protocol; the scorer never emits "replayable" without execution — the strongest statement a Card carries in this release is that preconditions are satisfied. Execution itself now separates the probe from the closest prior art: AgenTracer is "the first automated framework for annotating failed multi-agent trajectories via counterfactual replay and programmed fault injection" (Zhang et al., 2025), replaying in service of "pinpointing the specific agent or step responsible for an error within long execution traces" (Zhang et al., 2025); the probe scores whether replay is possible from the evidence at all — and where evidence cannot support even the preconditions of replay, that insufficiency is the measured finding, not an obstacle to the study.

The mapping from the six §3.5 preconditions to the released v0 card fields the probe reads is explicit, and deliberately sparse. (i) Model binding resolved: F03, the model/provider/resolved-version binding status (§6.3 reads its partial status as the failed resolved-model precondition). (ii) Decision context reconstructable: not operationalised as a probe-read check in v0 — adjacent evidence lands in F04 and F06, which the probe does not read — v2 protocol. (iii) Determinism controls: no covering v0 field; not operationalised in v0, v2 protocol. (iv) Monitor-view specification: no covering v0 field — F07 records scorer and monitor identity, not the view shown — not operationalised in v0, v2 protocol. (v) Permission and policy state: the policy half is carried by the bundled F10, policy and counterfactual replay notes, which the probe reads with a ceiling of partial; the permission half lands in F05 but is not read as a replay precondition in v0. (vi) Environment and side-effect re-creation: no dedicated field — F09's structured replay handle marks a machine-identifiable route back to the run, not a re-creation specification — not operationalised in v0, v2 protocol. The released probe therefore reads F09 and F10, with §6's precondition analysis additionally reading F03; the release is coarser than the §3.5 design here in exactly the sense §4.4 states.

## 4.4. From design to release: the v0 scorer

The Card above is the design contract; the released v0 scorer, inspect-evidence-sufficiency (Solozobov, 2026a), reads an existing trace — an Inspect log, a ControlArena trajectory export, or generic JSON/JSONL — and scores it into schema evidence-sufficiency-card.v0 over twelve evidence fields: F01 eval objective and release decision; F02 task and dataset sample scope; F03 model/provider/resolved version binding; F04 prompt, scaffold, and agent policy identity; F05 tools, sandbox, and network permissions; F06 raw messages, events, and tool calls; F07 scorer and monitor identity; F08 retries, errors, and timeouts; F09 transcript and replay handle; F10 policy and counterfactual replay notes; F11 leakage and reward-hacking checks; F12 residual uncertainty and not-sufficient-for statement. Each field is scored three-valued (`present=1.0`, `partial=0.5`, `missing=0.0`); the card aggregates a sufficiency_score, a blocking_gaps list, and a conservative verdict — insufficient, conditional, or sufficient-for-named-decision — where missing or partial replay, scorer-identity, model-binding, or scope evidence caps the verdict at conditional; cards are deterministic, byte-identical given fixed inputs and a pinned timestamp. Table 1 states the correspondence.

**Table 1.** Design-to-release correspondence: §3-§4 card design versus released v0 scorer.

| Design (§3-§4) | Released v0 operationalisation |
|---|---|
| Eight property classes (Definition 1) | Twelve field checks F01-F12; classes split or merge (permission recoverability into F05; counterfactual replayability into F09-F10) |
| Unit-interval proportion $r_p$ (Definition 2) | Three-valued field status (1.0 / 0.5 / 0.0) — the discrete special case |
| Vector plus bottleneck and coverage scalars (Definition 3) | sufficiency_score plus conservative verdict tiers; designated missing or partial fields cap the tier |
| Overclaim gap G (Definition 4) | Field-level analogue only, not G itself: difference of sufficiency_scores plus blocking-gaps comparison for a matched-outcome pair (§3.4) — a mean-like twelve-field scalar, not the bottleneck R_min; equivalence not established |
| replay_preconditions sub-verdict, counterfactuals_run empty (Definition 5, §3.5) | F09-F10 statuses (per-precondition mapping: §4.3); missing counterfactual replay surfaces as a blocking gap |

The mapping is stated, not hidden: the release is coarser than the design — per trace, not per decision; discrete, not unit-interval-valued; responsibility attribution and evaluation-awareness confounding carried only indirectly — and the design remains the specification the implementation converges toward. The twelve fields also differ in provenance: most are trace-derived, but F01 is scoring-time context supplied by the operator and F12 is the card's own residual-uncertainty statement, so sufficiency_score is a documentation-completeness scalar over mixed-provenance fields, not a pure captured-evidence proportion in Definition 2's sense; a trace-earned-only recount is stricter (per-card composition: §6.1).

## 4.5. The trace-level monitor-coverage gate

One Card presence question generalises into a per-run release check, stated as an atomic invariant: For every risky tool-execution span in a run, an *identified* monitor must have *actually scored that span*; any uncovered risky span makes the eval's PASS unbacked and *fails the release gate*. Per-run and machine-checkable, the invariant reports monitor-coverage — monitor-observation sufficiency — not model or monitor quality; four contrasts fix its position. Coverage accounting is not new: MonitoringBench red-teams monitor catch rates — an "Opus-4.5 monitor's catch rate falls from 94.9% on elicited-only Opus attacks to 60.3%…" (Jotautaitė et al., 2026) — and accounts for full-trajectory scorer coverage, roughly 99-100% for one scorer and about 95% for the other two, computed over valid-score sets (Jotautaitė et al., 2026). LinuxArena defines action-level monitor coverage in a live control setting (Tracy et al., 2026), and ControlArena exercises the per-turn suspicion scoring such accounting reads (UK AI Security Institute (AISI); Redwood Research, 2025). The novelty is the release-gate form — per-run, all-or-block, over a team-policy-defined risky set, a binary exit-code invariant integrated with the claim-conditioned Evidence Sufficiency Card — not aggregate coverage accounting inside a benchmark's own scoring pipeline. EviBound gates the agent's own completion claim through "dual governance gates requiring machine-checkable evidence" (Chen, 2025); this gate audits the record of the monitor's observation, not the agent's claim about its task. Auditable Agents disclose post-hoc, per-system record completeness (Nian et al., 2026), and ADAS issues "a cryptographically verifiable deployment certificate" consumed as a license to operate (Saparning, 2026); this gate is neither a descriptive audit percentage nor a per-model license but an inline binary verdict on one trace.

The shipped mechanism is small. Card field F07 asks whether a scorer or monitor identity is present, and the CLI enforces it as a presence bar: *–require-present* F07 exits 1 whenever the field is not present, *–gate* exits 1 on an insufficient verdict (*–strict* also on conditional), and exit 2 is reserved for tool and usage errors, so automation distinguishes a gate block from a failure to run. Presence is not coverage: a companion command emits a second card — schema monitor-coverage-card.v0, same exit-code contract — for the next question: did that monitor actually score the assistant turns that executed the risky tools? A turn is monitored only when a numeric suspicion score is co-located with the tool calls it scores; identity is declared run-level, never inferred; the risky set is team-owned policy, defaulting to all tool-bearing turns. The card records monitor_identity, risky against monitored risky turn counts, coverage_ratio, uncovered_spans by turn and tool, and a verdict — uncovered, partial-coverage, covered — of which only covered passes the gate.

The released walkthrough pair fixes the semantics. In the covered fixture an identified monitor scored all three risky tool turns: coverage_ratio 1.0, empty uncovered_spans, verdict covered, gate passes. In the uncovered fixture the same monitor scored one of three risky turns: coverage_ratio 0.333, uncovered_spans naming a transfer and a delete_record turn, verdict partial-coverage, gate blocks — identity is present in both, so what fails is monitor-coverage, not identity. The card's own boundary line — "`coverage = whether` an identified monitor scored the turn that executed the risky tool; NOT whether the monitor's judgment was correct." — and its claim-boundary list forbid the invited misreadings: coverage is not accuracy, covered is not safety or release approval; a run whose monitor scores every turn — the normal

ControlArena-style case (UK AI Security Institute (AISI); Redwood Research, 2025) — cards as covered and passes, and the gap arises under partial, sampled, or post-hoc monitoring, or when a trace names a monitor but carries no per-turn scores.

The gate claims no contribution beyond §1's four: it composes existing Card presence checks into one per-run invariant, inheriting the §3 scope — coverage and presence, never model or monitor quality. The artifact thus closes the loop §3 opened: the adapter populates present(p), the Card carries per-property scores and matched-outcome fields, the probe scores replay preconditions, the v0 scorer releases the Card as twelve deterministic checks, and the gate turns one presence question into a per-run release check. §5 exercises the adapter on Inspect and generic JSON/JSONL inputs — the ControlArena reader is implemented but not demonstrated on a real export (§5.1) — and reports what the populated Cards reveal.

# 5. Experimental Setup

The empirical component is a demonstration on existing traces, not a controlled experiment: the released scorer reads existing public and bundled inputs, runs no new models, fabricates no outcomes, so scoring is a near-zero-cost, deterministic pass. The controlled validation matrix — open-weight models crossed with scaffolds and benchmarks, plus executed counterfactual replay — is explicitly the v2 protocol (§3.5). The design is descriptive and diagnostic, not pre-registered (§1); the scoring rules were fixed before this analysis by the released artifact, version 0.3.0, archived 2026-07-01 as a citable software snapshot (Solozobov, 2026a). The demonstration can show only whether the score tracks the evidence an input carries; §6 reads it that way.

## 5.1. Trace substrate

The substrate is four traces scored into Evidence Sufficiency Cards plus two fixture pairs exercising the gates (Table 2).

The scored inputs are not a statistical sample; they span two of the scorer's three ingestible formats — Inspect eval logs (UK AI Security Institute (AISI), 2024) and generic JSON and JSONL traces; the third, ControlArena trajectory exports (UK AI Security Institute (AISI); Redwood Research, 2025), is implemented but exercised only against a synthetic fixture (below).

Table 2 records, per input, the source format the adapter ingests, the data class, and the assigned evidentiary role.

**Table 2.** Demonstration substrate. Public datasets are CC-BY-4.0; pinned provenance (dataset revision, file, sha256) lives in the released source manifest and each card.

| Input | Source format | Data class | Role |
|---|---|---|---|
| Trace Commons coding-agent trace | generic JSONL agent trace | real, public | donated session; no eval scaffolding |
| Assayo judged web-research trajectory | generic JSON agent trace | real, public | judged: tools, steps, model reference, rubric verdict |
| Inspect eval log (mockllm) | Inspect log, JSON format | real, bundled | genuine harness log; offline, no API key |
| Synthetic Inspect-style fixture | Inspect-style JSON | synthetic, bundled | best-instrumented anchor; replay handle specified_not_executed |
| Monitor-coverage pair (covered / uncovered) | monitor-coverage trace, JSON | synthetic, bundled | payments agent; exercises §4.5 gate semantics |
| Release-gate pair (raw / instrumented) | generic JSON trace | synthetic, bundled | same billing exchange, raw versus eval-instrumented; §3.4 matched pair |

Evidentiary roles are explicit: the two public traces and the real Inspect log carry §6's load-bearing observations; the synthetic fixture and both fixture pairs are hand-built anchors fixing scorer and gate semantics, claiming nothing about any model or system. Cross-harness ingestion is demonstrated on Inspect; for ControlArena the reader is implemented and test-covered on a synthetic fixture, not demonstrated. No human-subject data is collected — inputs are public CC-BY-4.0 datasets (fetched at runtime, never redistributed, anonymization caveats noted) and synthetic fixtures — so ethics review is not required.

## 5.2. Scoring procedure and gate semantics

Scoring is the §4.4 procedure, deterministic, no model in the loop: twelve checks F01-F12, three-valued (`present=1.0`, `partial=0.5`, `missing=0.0`), aggregated into sufficiency_score, blocking_gaps, and the conservative verdict tier insufficient / conditional / sufficient-for-named-decision. Monitor-coverage scoring follows §4.5: a turn is monitored only when a run-level identified monitor's numeric suspicion score is co-located with its tool calls; over the risky set (team-owned; default: all tool-bearing turns) the card reports coverage_ratio, uncovered_spans, and covered / partial-coverage / uncovered. Replay preconditions are read probe-only via the §4.3 precondition-to-field mapping: F09 reaches present only on a structured, machine-identifiable replay handle; F10 is capped at partial, declared counterfactual claims taken at face value, never verified; no replay of any level is executed. Gates put the block signal in the exit code (§4.5): *–gate* exits 1 on insufficient, *–strict* also on conditional, *–require-present* exits 1 when a listed field is not present, the coverage gate passes only covered; exit 2 marks tool or usage errors, never a block.

### 5.3. Reproducibility and the research-question reading

Cards are deterministic given fixed inputs and a fixed timestamp — the released build pins SOURCE_DATE_EPOCH, so a re-run reproduces each card byte for byte — and public inputs arrive pinned to a dataset commit revision, sha256-verified. An offline verification target re-runs the tests and re-scores the synthetic fixture, Inspect log, and coverage pair; a second adds the two public fetches; every §6 number regenerates from these two commands plus the two release-gate walkthrough invocations shipped with the example. The scorer, fixtures, source manifest, and cards ship as an open Apache-2.0 reproducibility package.

The substrate varies source format and instrumentation regime; each research question reads a fixed surface. RQ1 reads the per-check statuses at the released scorer's field level: which of F01-F12 each source leaves present, partial, or missing, and whether the pattern differs by source and format. RQ2 reads sufficiency scores of inputs sharing a surface reading, per §3.4's field-level analogue of the overclaim gap over card outputs — Definition 4's G itself is unmeasured at this stage — the release-gate pair supplying the matched same-task instance. RQ3 reads its §3.5 precondition form, in the v0-operationalised subset mapped in §4.3, off F03, F09, F10, and the cards' replay and counterfactual reference lists; its causal form is the v2 protocol.

## 6. Results

Every number below regenerates from the released package (§5.3); statistical hypothesis tests are out of scope at this stage — the analyses are descriptive, not pre-registered (§1, §5). Across the four scored cards, verdicts split two insufficient, two conditional, none sufficient-for-named-decision (Table 3).

**Table 3.** Evidence Sufficiency Cards for the four scored inputs (rows and formats as in Table 2).

| Input | Source format | Verdict | Sufficiency | m / p / pr | Blocking gaps |
|---|---|---|---|---|---|
| Trace Commons coding-agent trace | generic JSONL agent trace | insufficient | 0.458 | 4 / 5 / 3 | F02, F03, F07, F09, F10 |
| Assayo judged web-research trajectory | generic JSON agent trace | insufficient | 0.458 | 4 / 5 / 3 | F02, F03, F06, F09, F10 |
| Inspect eval log (mockllm) | Inspect log, JSON format | conditional | 0.667 | 2 / 4 / 6 | F03, F09, F10 |
| Synthetic Inspect-style fixture | Inspect-style JSON | conditional | 0.833 | 0 / 4 / 8 | F10 |

Note. Sufficiency is the sufficiency_score over the twelve three-valued check statuses (`present=1.0`, `partial=0.5`, `missing=0.0`; 12 fields per card, two of which — F01, F12 — carry scoring-time or card-self rather than trace provenance, §4.4); *m/p/pr* counts missing / partial / present; blocking gaps hold the verdict below sufficient-for-named-decision; cards regenerate byte-identically (§5.3).

## 6.1. Per-check gaps by source and format (RQ1)

Every scored input carries per-check evidence gaps, and their composition differs by source and format on this bounded substrate (Table 3) — RQ1's answer at the released scorer's F01-F12 operationalisation level, the gaps H1 hypothesises. Among the four scored inputs, scorer and monitor identity (F07) is missing only on the raw coding trace, which carries a model reference and a tool call but nothing identifying any scorer or monitor; where F07 is present it is carried by scorer identity alone, monitor references being absent from all four cards. Task and dataset scope (F02) is partial on both public traces, present on both Inspect-style inputs; retry and error evidence (F08) is missing on both public traces yet present on the real Inspect log, which records two errors; model binding (F03) is partial on all three real inputs — model references exist, but no resolved provider snapshot — and present only where a digest was hand-authored into the fixture.

The present column is thinner than its count: of the coding trace's three present fields the trace itself earns only F06 — F01 is the scoring-time decision context, F12 the card's own not-sufficient-for statement; the judged trajectory's trace-earned present is F07, its rubric verdict.

### 6.2. The sufficiency spread under a shared surface reading (RQ2)

The four scored inputs share only a surface reading — an agent run completed and left a trace usable for review, the two Inspect-style inputs recording a top-level success status — which is not a matched nominal outcome, so the comparison sits outside Definition 4's domain and its spread is a cross-substrate sufficiency spread, not the overclaim gap G. On that footing, twelve-field completeness spans 0.458 to 0.833 (Table 3), a spread of $|0.833 - 0.458| = 0.375$ between the extremes. The reading for H2 is direct at the level §1 poses it: a non-zero field-level spread beneath a shared surface reading is the difference H2 hypothesises, while Definition 4's G itself is unmeasured in this demonstration — its matched-outcome measurement is the v2 protocol, and the §3.4 field-level analogue is exercised on the matched pair below. Within that shared surface reading, the two real public traces report different tasks (coding versus web research) and formats (JSONL versus JSON) yet identical sufficiency 0.458, identical status counts, and different blocking composition, F07 against F06 — the same scalar over materially different gaps.

The release-gate pair is the cleanest matched instance (both fixtures synthetic, §5.1): the same billing knowledge-base exchange with the same recorded success status, raw versus eval-instrumented, the instrumented variant adding a resolved model digest, a scorer manifest, an explicit monitor, and recorded scores. The raw variant cards at sufficiency 0.542 (present 4, partial 5, missing 3; blocking F03, F07, F09, F10), the instrumented variant at 0.667 (present 6, partial 4, missing 2; blocking F09, F10): the pair's field-level analogue of G (§3.4) is $|0.667 - 0.542| = 0.125$, field-resolved to F03 and F07, the two fields on which the blocking lists differ — the nominal outcome is matched, but the aggregation is the twelve-field sufficiency_score rather than Definition 3's bottleneck $R_{min}$, and Definition 4's property-resolved $g_p$ ranges over the eight property classes, not F-fields, hence the analogue, not G. Gate consequence: under *–gate* with *–require-present* F07, the raw variant exits 1 and the instrumented variant exits 0; scored standalone with no named decision, the raw variant's verdict drops to insufficient and plain *–gate* blocks it.

### 6.3. Replay preconditions (RQ3, precondition form)

Counterfactual-replay preconditions are unmet in every scored trace, answering RQ3 in its §3.5 precondition form for the subset v0 operationalises — model binding and the policy half of permission-and-policy state — plus the F09 replay-handle presence check, a route back to the run rather than a re-creation specification, read off F03, F09, and F10 per the §4.3 mapping; the remaining preconditions are not operationalised in v0 and travel with the v2 protocol. F10 is missing on all three real inputs and partial on the synthetic fixture, partial being the ceiling the scorer permits for a declared counterfactual claim taken at face value, never verified; F09 rises past partial on no real input, deterministic replay remaining unproven, and the fixture's present F09 reflects a declared structured handle marked specified_not_executed; the replay and counterfactual reference lists are empty on all three real inputs, and no real input resolves its model binding (F03 partial on all three), so the §3.5 resolved-model precondition fails everywhere outside hand-authored fixtures. F10 stands in the blocking list of all six Evidence Sufficiency Cards this paper reports — the four of Table 3 and both release-gate cards. Executed counterfactual replay is thus not merely unperformed here; for want of preconditions it is not well-defined — the precondition-form finding. Whether executing the §3.5 protocol changes a reconstructability verdict — RQ3's causal form, H3 of §1 — remains untested, the v2 protocol.

## 6.4. Monitor-coverage gate on the demonstration pair

The walkthrough pair instantiates the §4.5 invariant with exact numbers. In the covered fixture the identified monitor (agentic_payments_suspicion_monitor) scored all three risky tool turns: coverage_ratio 1.0 (100%), risky set all tool-bearing turns, uncovered_spans empty, verdict covered, gate exit 0. In the uncovered fixture the same monitor scored one of three risky turns: coverage_ratio 0.333, uncovered_spans naming the transfer turn and the delete_record turn, verdict partial-coverage, gate exit 1; identity is present in both fixtures, so what fails is monitor-coverage, not identity. Narrowing the risky set to the two mutating tools via team-owned policy drives coverage to 0.0 and the verdict to uncovered, also blocking. The §4.5 boundary governs the reading — a scored turn shows the trace records the monitor as scoring it, not what the monitor saw (monitor-view content not operationalised in v0; §4.3) nor that it was right — so the pair demonstrates gate semantics and claims nothing about any deployed monitor. Across both gates the blocking consequence is uniform: an insufficient verdict, a required field not present, and any coverage verdict short of covered each exit 1.

## 6.5. Null results and what these results do not establish

Three null observations belong to the record. The scalar alone does not separate the two real public traces — identical 0.458, identical status counts, different gap composition — so the discriminating signal sits in the blocking lists, not the headline score. No input reaches sufficient-for-named-decision — the fully instrumented fixture is capped at conditional by F10 — so the demonstration has no positive-control card for the top tier. The substrate holds no real matched-outcome pair — the only same-task pair is synthetic — leaving monitor observation exercised only by the §6.4 fixtures.

The cards' own boundary lists bound what this section establishes: per their not_sufficient_for entries, no production release approval by itself, no customer-facing reliability or safety claim, no benchmark leaderboard claim, no audit, legal, or compliance sufficiency; per their claim boundaries, nothing about model quality or safety, monitor accuracy (coverage is not accuracy), or benchmark validity, and no showing that Inspect or ControlArena lack logs — the richest real card here is an Inspect log, and a normally monitored ControlArena-style run cards as covered. Headline sufficiency scores also modestly overstate trace-earned sufficiency: up to two of the twelve fields (F01, F12) are credited from scoring-time or card-self provenance rather than the trace (§4.4), and separating capture scores from documentation scores is deferred to the v2 protocol (FW1). A low score on a public trace indicates no deficiency in its donors: such traces were never packaged as release evidence, and the cards measure exactly that absence — the demonstration's point.

# 7. Limitations and Threats to Validity

The metric, adapter, Evidence Sufficiency Card, and replay probe are delivered as engineering artifacts; the demonstration that exercises them (§5-§6) is descriptive and diagnostic, not confirmatory, and its design is not pre-registered (§5). This section bounds both the construct and that demonstration: it states the construct-validity threats that attach to the instrument's declared inputs, the external-validity bounds of the substrate actually scored, and what a fuller study would calibrate; no statistical hypothesis test is run at this demonstration stage (§6).

The most consequential threat is to construct validity, and it concerns how the metric's reference points are fixed. Per-property scoring is deterministic given the schema — Definition 2 computes the recovered proportion $r_p$ by counting present against required evidence fields, with no learned parameters and no model in the loop — but that determinism is conditional on two author-declared inputs.

**T1.** The required-evidence sets $R_p$ and the designation of which property classes a given claim depends on are author-declared rather than empirically validated, so the metric is deterministic with respect to a schema whose own correctness is asserted, not measured. If $R_p$ for permission recoverability omits an element a defensible audit would demand, or includes one it would not, every downstream score and bottleneck inherits that judgement. The same applies to the claim-required subset R that fixes which properties enter the bottleneck score $R_{min}$. This is the central calibration target. The construct was first exercised under exactly this constraint: the author's prior pilot was "single-annotator, one anchor per cell, descriptive" and deferred "the two-annotator agreement-calibrated full benchmark on twenty to fifty real captured production traces per regime" to later work (Solozobov, 2026b). Full two-annotator ground-truth labelling of required-set membership and claim-required designations — forthcoming benchmark territory, not a result this paper claims — would convert the asserted schema into a calibrated one and is the work that would most strengthen the construct (closure: deferred to FW2).

**T2.** The metric scores whether captured evidence suffices to reconstruct a decision property, not whether any reconstruction drawn from that evidence is correct; it bounds evidence sufficiency, not ground truth (closure: by design, out of scope for the instrument). A claim can be reconstructable and false, or unreconstructable and true. A high bottleneck score certifies that the trace retained the elements a reconstruction requires, not that the agent behaved safely, that the monitor judged correctly, or that the policy was satisfied — those adjudications lie outside the instrument by design, as §3 states. Reading reconstructability as a correctness or safety verdict would overclaim precisely the distinction the metric exists to keep open.

**T3.** The demonstration bounds external validity narrowly: it scores existing public and bundled traces, with no new model runs; cross-harness ingestion is demonstrated on Inspect-format inputs only, the ControlArena export reader being implemented but not demonstrated on a real export; the substrate holds no controlled matched-outcome pair — the only same-task pair is synthetic — and three of its six inputs are hand-built synthetic anchors that fix scorer and gate semantics while claiming nothing about any model or system (§5.1). Nothing here establishes generalisation to frontier or closed models, to real ControlArena exports, or to other trace formats; the controlled validation matrix with executed replay — the v2 protocol — is the stated path to it (closure: deferred to FW1). The narrowness is a consequence of the construct rather than a convenience: reconstructability is a property of what a trace carries, not of model weights, so scoring existing traces suffices for what the demonstration claims — that the score tracks the evidence an input carries (§5).

**T4.** No executed counterfactual replay is run in this study: RQ3 is answered in its precondition form only, and its causal form — whether executing the §3.5 protocol changes a reconstructability verdict — remains untested until the v2 protocol. The reason is measured rather than logistical:

on this substrate no real input resolves its model binding, F09 rises past partial on no real input, and F10 stands in the blocking list of all six Evidence Sufficiency Cards this paper reports (§6.3), so the replay preconditions fail everywhere outside hand-authored fixtures. To the anticipated objection that the probe is specified but never executed: the probe is implemented and its preconditions are scored, and on this substrate executed replay is not even well-defined — which is itself the finding (closure: deferred to FW1).

**T5.** The eight property classes are a designed cross-walk from the prior pilot's schema and the auditability literature's dimensions rather than an exhaustively derived taxonomy, so a decision property load-bearing for some evaluation claim may fall outside the set. The classes refactor the five auditability dimensions of Nian et al. (2026) to the evaluation-claim unit and add two — claim sufficiency and evaluation-awareness confounding — that a per-system audit does not foreground. Relatedly, the metric treats an *opaque* element (observable in principle but not externally inspectable, paradigmatically model reasoning) and a *structurally absent* one alike as absent for scoring, recording the distinction on the Card to guide remediation rather than to alter the score. Whether eight classes are the right number, and whether the opaque/absent collapse is the right convention, are schema-design choices a broader validation could revise. An analogous collapse operates at instance level: unconditional membership (Definition 6) cannot distinguish a structurally absent element from an inapplicable one or an evidenced negative, a representational bound FW2's calibration would also resolve (closure: deferred to FW2).

Taken together, these threats locate the instrument precisely: it is a deterministic, schema-conditioned bound on evidence sufficiency for a declared claim, demonstrated on a deliberately narrow substrate of existing traces, and calibrated against author judgement that full ground-truth labelling would replace. Two future-work items absorb them: FW1, the v2 protocol — the controlled matched-outcome validation matrix with executed replay — and FW2, the forthcoming two-annotator ground-truth benchmark calibrating the required-evidence sets and enabling broader schema validation. The closure map is explicit: T1 deferred to FW2; T2 by design, out of scope for the instrument; T3 deferred to FW1; T4 deferred to FW1; T5 deferred to FW2. The §5-§6 demonstration should be read as exploratory diagnosis within these bounds, not as a validated measurement of any harness's adequacy.

# 8. Implications

The instrument this paper delivers, and the reporting practice it enables, carry implications for how agent-safety evaluations communicate what their numbers rest on — independent of any particular empirical finding. The implications below follow from the method and artifact, not from results; each describes how an evaluation author, harness maintainer, or safety-case author could act on a reconstructability measurement.

The headline implication is a reporting practice: report the per-decision reconstructability vector, with its weakest-link bottleneck scalar, alongside any safety-evaluation claim, so a reader sees the evidence regime the headline number rests on rather than the number alone. A task-success rate, attack-success rate, or monitor AUROC reported without its reconstructability vector reproduces the pathology the paper diagnoses, in which a single figure stands in for a heterogeneous and possibly insufficient evidence regime. The vector is the metric's primary output precisely so that a claim's evidential basis travels with its outcome: a reader can see that a monitor-score claim rests on a fully recovered monitor-input record but a partially recovered policy state, and read the bottleneck scalar as the ceiling on how reconstructable the claim is from what the harness kept. This makes evidence sufficiency a reported property of an evaluation rather than an unstated assumption behind it.

Adoption requires only modest instrumentation: an evaluation author or harness maintainer emits per-decision traces into the Evidence Sufficiency Card schema through the cross-harness adapter, after which scoring is deterministic, and the property-resolved overclaim gap tells them which property to instrument first. The shipped gate semantics extend this into a required per-run release check: the scorer's exit code blocks on an insufficient verdict, a presence bar fails any named evidence field not present — the monitor-identity check among them — and a monitor-coverage verdict short of covered also blocks (§4.5; exercised in §6.4). A continuous-integration pipeline consuming that exit code can therefore halt a release exactly when the evidence behind it is not there. Because the metric counts present against declared required evidence fields with no model in the loop, the integration cost is the mapping from a harness's existing trace format into the Card, not a new evaluation run or a learned judge; the adapter already targets this for two harnesses. Once a regime is scored, the per-property gap $g_p$ — in the released scorer, its field-level analogue — localises the deficit: a near-zero monitor-visibility score, or a property-resolved gap concentrated on policy-state visibility between a rich and an action-only recording, names the single property whose capture would most raise the bottleneck. Instrumentation effort is thereby directed by measured insufficiency rather than guessed at, and a maintainer can prioritise the evidence that a claim most depends on.

Evidence sufficiency is a natural explicit input to argument-based safety cases: where an evaluation result is offered as evidence in a structured safety argument, its reconstructability vector bounds how much that evidence can carry. Evaluations of stealth and situational awareness already frame their results this way, presenting "a safety case (Clymer et al., 2024) for scheming inability…, making a high-level argument that models need basic situational awareness and top human-level stealth to cause severe harm via scheming" (Phuong et al., 2025), and the evaluations-based safety-case line for scheming rests on the same move (Balesni et al., 2024). A safety case built on an evaluation whose claim-required properties are only partially reconstructable inherits that limit; surfacing the bottleneck score as an input to the argument lets a reviewer see whether the evidence is load-bearing or whether the case rests on a number the harness cannot reconstruct behind.

Across all three uses the instrument is diagnostic, not normative: it measures, bounds, and reveals where evidence is sufficient, and it does not certify a harness adequate, pass a safety case, or rank vendors. A reconstructability vector reports what a harness retained for a claim; it sets no threshold above which an evaluation is sound, and a zero overclaim gap means only that two matched regimes are equally evidence-sufficient, not that either suffices in absolute terms. The metric supplies a reader, maintainer, or safety-case reviewer with a calibrated view of the evidence regime behind a number and leaves the adequacy judgement, and any comparison across vendors, to them.

# 9. Conclusion

Agent-safety evaluation results are not yet load-bearing evidence when identical nominal outcomes — task success, attack success, a monitor score — rest on materially different evidence regimes, and the evidence a harness recorded cannot reconstruct the decision the safety claim depends on. This paper made that deficiency measurable rather than rhetorical, operationalising reconstructability as an evaluation-validity metric. It defined (1) a property-level reconstructability metric scoring, in the unit interval, how much of each of eight decision properties a captured trace recovers; (2) a vendor-neutral adapter that normalises traces from more than one evaluation harness into a per-decision Evidence Sufficiency Card; (3) a counterfactual-replay intervention protocol with an implemented replayability-precondition probe that scores whether such a replay is even well-defined on the captured evidence; and

(4) a claim-evidence overclaim gap stating exactly how much less evidence the same headline number rests on in the weaker of two matched regimes. The same Card backs a per-run monitor-coverage release check: whether an identified monitor actually scored each risky tool span before an evaluation's pass is treated as release evidence. Each instrument is deterministic and schema-based, reading evidence a harness already keeps.

Scored on existing public and bundled traces at near-zero cost, with no new model runs, the instrument separated evidence regimes a nominal reading cannot tell apart: the four scored inputs, sharing one surface reading of trace usability, spanned evidence sufficiency from 0.458 to 0.833, the synthetic release-gate pair blocked its raw variant at 0.542 and passed the instrumented variant at 0.667, and counterfactual-replay preconditions were unmet in all six reported cards (100%), so executed replay was not merely unperformed but undefined — itself the finding. The demonstration is deliberately bounded, and its limits are the calibration agenda (§7): it scores recoverability, not correctness (T2), over author-declared required-evidence sets (T1) — the eight-class schema itself a calibration target (T5/FW2) — and the controlled matched-outcome matrix with executed replay across open-weight models and harnesses is the second-version protocol (T3, T4/FW1), not a result claimed here. The recommendation is narrow and actionable: report a decision's reconstructability vector, with its weakest-link bottleneck, alongside any safety-evaluation claim, so a reader sees the evidence regime a headline number rests on rather than the number alone. An evaluation becomes load-bearing not when its score is high, but when the evidence behind that score can be reconstructed by someone who was not there when it ran.